\documentclass[prl,aps,twocolumn,superscriptaddress,floatfix]{revtex4}
\usepackage{amssymb,amsmath,array,graphicx,epsfig}

\usepackage{subfigure}

\usepackage{multirow}

\begin{document}

\title{A defect mediated lamellar to isotropic transition of amphiphile bilayers}

\author{Antara Pal}
\affiliation{Raman Research Institute, C V Raman Avenue, Bangalore 560 080, India}
\author{Georg Pabst}
\affiliation{Institute of Biophysics and Nanosystems Research, Austrian Academy of Sciences, 8042 Graz, Austria}
\author{V. A. Raghunathan}
\affiliation{Raman Research Institute, C V Raman Avenue, Bangalore 560 080, India}


\begin{abstract}

We report the observation of a novel isotropic phase of amphiphile bilayers in a mixed system consisting of the ionic surfactant, sodium docecylsulphate (SDS), and the organic salt p-toludine hydrochloride (PTHC). This system forms a collapsed lamellar ($L_\alpha$) phase over a wide range of water content, which transforms into an isotropic phase on heating. This transition is not observed in samples without excess water, where the $L_\alpha$ phase is stable at higher temperatures. Our observations indicate that the $L_\alpha$ - isotropic transition is driven by the unbinding of edge dislocation loops  and that the isotropic phase in the present attraction-dominated system is the analogue of the sponge phase usually seen in amphiphile systems dominated by interbilayer steric repulsion.


\end{abstract}

\maketitle


Amphiphiles self-assemble in water to form aggregates above a critical micellar concentration (CMC). The shape of these aggregates is in turn determined by that of the amphiphile, which is usually described in terms of the shape parameter $p=v/al$, where $v$ is the volume of the hydrocarbon chain of the amphiphile, $a$ the optimal head group area and $l$ the typical length of the chain in the aggregate \cite{israel}. Bilayers are formed in dilute solutions by amphiphiles with $p \sim 1$, such as double-tailed phospholipids which are the major constituents of biomembranes. It is also the preferred morphology in mixtures of single-tailed cationic and anionic amphiphiles. In these so-called catanionic systems bilayers in the form of unilamellar vesilces (ULV) are present in dilute solutions near the equimolar composition of the two amphiphiles \cite{kaler89}. Organic salts, such as PTHC, can be considered as an extreme limit of an ionic amphiphile. While they do not self-assemble in water to form micelles, they have a preference to sit at the surface of the micelle and hence affect the properties of the micelle-water interface \cite{hassan02}. Such molecules, often called \emph{hydrotropes}, decrease the spontaneous curvature of the micellar surface leading to the formation of long worm-like micelles (WLM) in many amphiphiles, which in their absence aggregate into short cylindrical micelles \cite{porte89}. In some cases, the decrease in the spontaneous curvature is large enough to stabilize bilayers \cite{sajal07}.

The most common phase formed by bilayers is the lamellar phase consisting of a periodic stack of bilayers separated by water \cite{roux}. In the absence of any osmotic pressure, the lamellar periodicity of these systems is determined by the balance between different inter bilayer interactions, namely, electrostatic repulsion, van der Waal's attraction, steric repulsion and hydration repulsion \cite{israel}. Steric repulsion arises from thermal undulations of the bilayers \cite{helfrich78a}, whereas the short-ranged hydration force arises from the ordering of water molecules at the bilayer surface and from the protrusion of molecules out of the bilayer.   Bilayers also exhibit isotropic phases without any long-range positional or orientational order, such as a dispersion of ULV and the sponge ($L_3$) phase \cite{roux92}. Although ULV dispersions can be formed in many amphiphilic systems, only in catanionic systems they are believed to be truly thermodynamically stable \cite{kaler89,safran90}. The sponge phase, made up of a random network of bilayers, is usually found in some very dilute amphiphilic systems in the presence of a cosurfactant, such as a short-chain alcohol, and an inorganic salt, such as NaCl \cite{porte89a,gazeau89,skouri91a}. In these systems electrostatic interactions are screened out by the salt and the bilayer flexibility is enhanced by the cosurfactant; as a result inter bilayer interactions are dominated by steric repulsion.

In this paper we report a novel isotropic ($L_x$) phase of bilayers without long-range positional and orientational order, exhibited by the anionic amphiphile SDS in the presence of the organic salt PTHC. This system exhibits a $L_\alpha$ phase near equimolar concentration of the two species \cite{sajal09}. This phase exhibits limited swelling and as a result, a dispersion of multilamellar vesicles (MLV) is formed over a wide range of water content, which reversibly transforms into the $L_x$ phase on heating. Ionic conductivity and SAXS studies indicate that this transition is driven by the formation of pores or passages in the bilayer. A similar defect-mediated transition has been theoretically studied in the context of the smectic A - nematic liquid crystal transition \cite{helfrich78,nelson81,holyst94}. Our observations suggest that the $L_x$ phase is the analogue of the sponge phase in systems where van der Waal's attraction is the dominant inter bilayer interaction.

SDS (99\% purity) and PTHC (98\% purity) were obtained from Sigma-Aldrich and were used without further purification. Small angle X-ray scattering (SAXS) studies on unoriented samples were performed using a Hecus S3-Micro System, equipped with a one-dimensional position sensitive detector (PSD). SAXS data were analyzed using the procedure described in ref. \cite{pabst00}, the details of which are given in the supplementary material. Phase transitions in the samples were also monitored using polarizing optical microscopy. $\Phi_s$ denotes the combined weight percent of SDS and PTHC, whereas $\alpha$ is the PTHC to SDS molar ratio.


\begin{figure}[h]
\begin{center}
\includegraphics[width=70mm,angle=0] {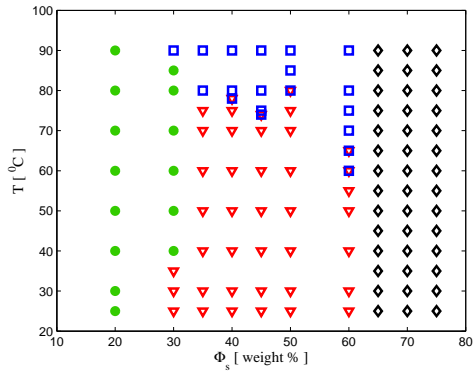}
\end{center}
\caption{T - $\Phi_s$ phase diagram of the SDS-PTHC-water system at $\alpha$ = 1.0. The symbols $\bullet$, $\triangledown$ , $\square$ and $\diamondsuit$ represent UB, MLV ($L_\alpha$ + excess water), $L_x$ + excess water and $L_\alpha$, respectively.}
\label{phasedia-alpha}
\end{figure}

\begin{figure}[h!]
\centering

\subfigure[]{
   \includegraphics[width=70mm,angle=0] {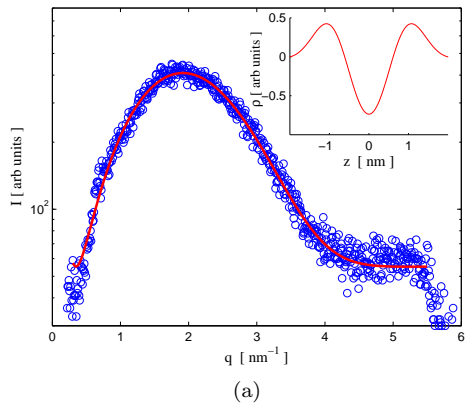}
   \label{saxs-ulv}
 }

 \subfigure[]{
   \includegraphics[width=70mm,angle=0] {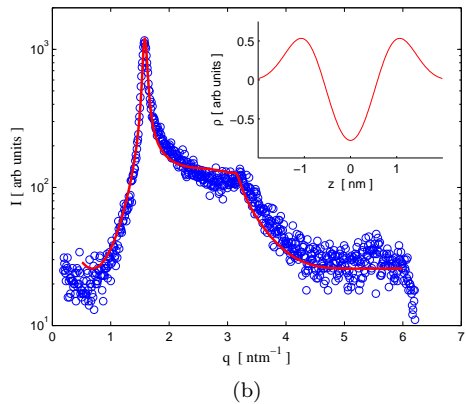}
  \label{saxs-mlv}
 }

 \subfigure[]{
  \includegraphics[width=70mm,angle=0] {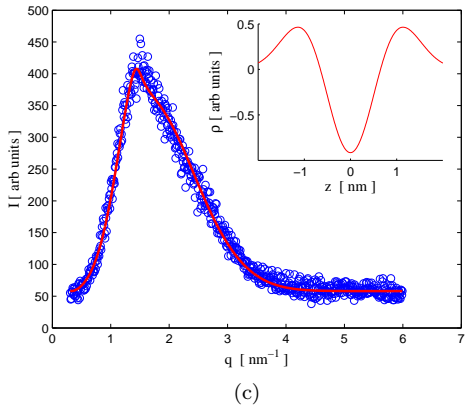}
 \label{saxs-lx}
 }

\caption{SAXS patterns of the different phases of the SDS-PTHC-water system. (a) UB at $\alpha=1.0$;$\Phi_s$ = 20, T = 60$^\circ$C (b) $L_\alpha$ at $\alpha=1.4$;$\Phi_s$ = 20, T = 10$^\circ$C  and (c) $L_x$ at $\alpha$=1.0;$\Phi_s$ = 30, T = 90$^\circ$C. The solid lines are fits obtained from the model (see supporting material). The insets show the corresponding bilayer electron density profiles. }

\label{saxs}
\end{figure}

Figure \ref{phasedia-alpha} shows the temperature - $\Phi_s$ phase diagram of the system at $\alpha$ = 1.0. At low surfactant concentrations the samples are slightly translucent and are isotropic under a polarizing microscope. Their SAXS patterns show a rather smooth variation of the scattered intensity with $q$, the magnitude of the scattering vector (fig. \ref{saxs-ulv}). These observations suggest the formation of a dispersion uncorrelated bilayers (UB) at these concentrations, as confirmed by  detailed analysis of the SAXS data (see supplementary material). However, from our data we cannot conclude whether these are unilamellar vesicles (ULV) or multilamellar vesicles (MLV) with uncorrelated bilayers.  With increasing $\Phi_s$, the $L_\alpha$ phase is formed as indicated by the appearance of two peaks in the SAXS pattern, with their spacings in the ratio 1:1/2 (fig. \ref{saxs-mlv}). The lamellar periodicity of this phase is $\sim$ 5.0 nm and it forms a turbid MLV dispersion as confirmed by polarizing optical microscopy observations. At $\Phi_s$ = 30 an \emph{unbinding transition} \cite{lipowsky86} is observed at around 40 $^\circ$C on heating, above which diffraction patterns show uncorrelated bilayers. Concomitantly, the samples lose their birefringence. This transition is reversible and the MLV form immediately on cooling. This indicates that above the unbinding transition the sample consists of MLV with uncorrelated bilayers and not ULV.  On increasing the temperature further,  oil-like droplets start phase separating out of the solution at $\sim$ 90$^\circ$C. These droplets  coalesce into a distinct lower layer with a well-defined meniscus separating it from the clear solution on top. From polarizing microscopy it is found that this optically isotropic phase exhibits flow birefringence under shear. SAXS patterns of this phase show a broad peak (fig. \ref{saxs-lx}), with an average spacing of around 4.0 nm which is very close to the periodicity of the lamellar phase occurring at lower temperatures. The $L_\alpha$ $\rightarrow$ UB $\rightarrow$ $L_x$ phase sequence is reversible on lowering the temperature.

For 35 $\leq$ $\Phi_s$ $\leq$ 60 we do not observe a dispersion of UB intervening between $L_\alpha$ and $L_x$, but SAXS data indicate the presence of increasing concentration of UB with increasing temperature near the transition to the $L_X$ phase for $35 < \Phi_s < 50$. The transition temperature gradually decreases with increasing $\Phi_s$ from 90 $^\circ$C at  $\Phi_s = 30$ to 70 $^\circ$C at $\Phi_s = 60$, with a narrow range of coexistence.  
For $\Phi_s \geq 65$ the samples contain no excess water, and the lamellar periodicity of the $L_\alpha$ phase decreases with increasing $\Phi_s$. Interestingly, these samples do not transform into the $L_x$ phase on heating up to 90 $^\circ$C. 

Within the narrow composition range where the pure $L_\alpha$ phase exists the variation of the d-spacing follows the swelling behavior given  by $ d = \delta /\phi $, where $\delta$ is the bilayer thickness and $\phi$ the amphiphile volume fraction, with $\delta$ = 2.49 nm (see supplementary material). The limited swelling of the $L_\alpha$ phase can be attributed to negligible long-range repulsive inter bilayer interactions in the system. Electrostatic repulsion can be expected to be weak due to the neutralization of the bilayer charge by the strongly binding aromatic counterion, and screening by the high concentration of NaCl in the solution. Absence of significant steric repulsion indicates a high bilayer bending rigidity ($\kappa$) of the order of 10 $k_B$T \cite{helfrich78a}. The attractive van der Waal's interaction under these conditions is balanced mainly by the short-range hydration repulsion.

\begin{figure}[h]
\begin{center}
\includegraphics[width=70mm,angle=0] {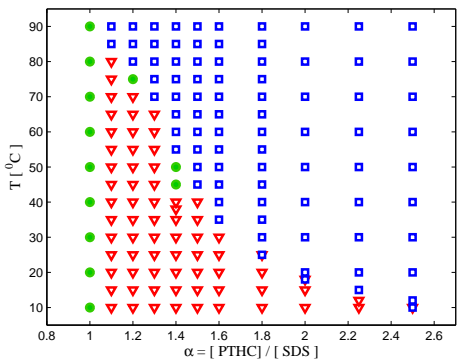}
\end{center}
\caption{ T - $\alpha$ phase diagram of the SDS-PTHC-water system at $\Phi_s$ = 20. The symbols $\bullet$, $\triangledown$, and $\square$ represent UB, MLV ($L_\alpha$ + excess water) and $L_x$ + excess water, respectively.}
\label{phasedia-phi}
\end{figure}

Figure \ref{phasedia-phi} shows the phase diagram of the system in the $\alpha$-T plane at $\Phi_s$ = 20. UB are obtained at lower values of $\alpha$ and with increasing salt concentration positional correlations between the bilayers increase, resulting in a $L_\alpha$ phase coexisting with excess solvent. At still higher values of $\alpha$ the $L_x$ phase is observed. At lower $\alpha$ the $L_\alpha$ phase in excess water forms a translucent MLV dispersion, whereas at higher $\alpha$ it phase separates completely forming a rough, but well-defined, meniscus separating it from the aqueous solution. On the other hand, the meniscus is always smooth in the case of the $L_x$ phase, consistent with its much lower viscosity. The broad peak in its SAXS pattern has an average spacing of about 4.0 nm and is rather insensitive to temperature and salt content.

\begin{figure}[h]
\begin{center}
\includegraphics[width=70mm,angle=0] {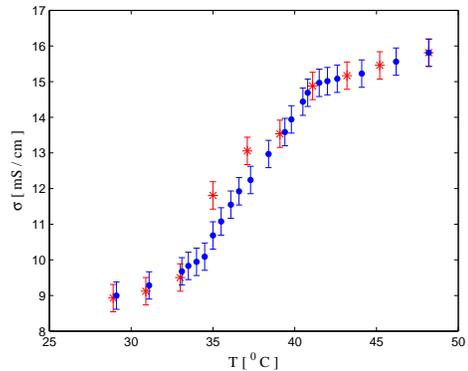}
\end{center}
\caption{Variation of the ionic conductivity of a SDS-PTHC-water sample ($\alpha$ = 1.5, $\Phi_s$ = 20) on heating ($\bullet$) and cooling ($\ast$) across the $L_\alpha$ - $L_x$ transition.}
\label{conduct}
\end{figure}

We have tried to fit the SAXS data to the scattering expected from the sponge phase \cite{roux92,lei97}. But it was not possible to get a good fit for any reasonable values of the model parameters. Further, the mesh size in the sponge phase (given by the position of the broad peak) is typically 1.5 times the lamellar periodicity of the $L_\alpha$ phase with the same water content \cite{skouri91}. However, in the present case the two spacings are comparable. Hence we can rule out the $L_x$ phase being a conventional sponge phase. Another possibility is that it is made up of inverted micelles. But it can be ruled out on the basis of the phase diagram (fig. \ref{phasedia-alpha}), which shows a pure $L_\alpha$ phase occurring at lower water content than the $L_x$ phase.

The ionic conductivity of a sample exhibiting the MLV $\rightarrow$ $L_x$ transition was measured as a function of temperature (fig.\ref{conduct}). These experiments were carried out after removing the coexisting aqueous solution in the lower temperature phase. The observed variation of the conductivity is very similar to that seen across the $L_\alpha$ - sponge transition in other systems \cite{hecht95}, where it has been interpreted in terms of formation of passages between the bilayers. 

Passages or pores in the bilayer are edge dislocation loops in the lamellar stack. The influence of such dislocation loops on the structure of a lamellar phase has been studied theoretically in the context of the smectic A - nematic liquid crystal transition \cite{helfrich78,nelson81,holyst94}. It has been shown that a finite density of unbound dislocations can destroy the (quasi) long-range positional order of the lamellar phase and make the interlamellar correlations short-range, as observed in the present case \cite{foot}. In the absence of excess water the creation of dislocation loops would lead to a decrease in the spacing of the stack far from the defect, since the material removed to create the defect has to be accommodated elsewhere. This will be resisted by the exponentially varying hydration repulsion between the bilayers. As a result such defects will be suppressed when there is no excess water. This situation is similar to the suppression of bilayer curvature defects with decreasing water content seen in some lamellar phases \cite{holmes93}, and the theoretically predicted suppression of dislocation loops in smectic films confined between two rigid boundaries \cite{holyst94}.
  
The deformation energy of a bilayer is described in terms of two elastic moduli; the modulus of mean curvature $\kappa$ and the modulus of Gaussian curvature $\bar{\kappa}$ \cite{helfrich73}. Theoretically it has been shown that the sponge phase consisting of a network of bilayers should be stable when $\bar{\kappa}$ is weakly negative; for lower values of  $\bar{\kappa}$ the lamellar phase is stable \cite{golubovic94,morse94}. In the case of bilayers made up of ionic amphiphiles there is an electrostatic contribution to $\bar{\kappa}$ which is predicted to be negative \cite{winterhalter88,mitchell89}. In the presence of a high concentration of salt the electrostatic interactions will be screened out and $\bar{\kappa}$ can become weakly negative, favoring bilayer passages and hence stabilizing the sponge phase, as observed in many systems.

Such passages can also be expected to form in the present system due to the high concentration of salt. However, the main difference between the present system and the ones where a sponge phase occurs is the difference in the inter bilayer interactions. The present system is dominated by attractive van der Waals interaction leading to a collapsed phase, whereas the sponge phase is seen in systems dominated by steric repulsion. The observation of a similar $L_\alpha$ - $L_x$ transition in some other amphiphile - water systems dominated by attractive inter bilayer interactions lead us to believe that the $L_x$ phase is the analogue of the sponge phase in such systems \cite{apal11}.     

In conclusion, we have observed a novel isotropic phase of amphiphile bilayers. This phase seems to be the analogue of the sponge phase, seen in swollen systems dominated by steric repulsion, in the van der Waal's attraction dominated regime. In both cases the transition from the lamellar phase is driven by the proliferation of bilayer passages. Although theoretical studies indicate that (quasi) long-range positional order of the lamellar phase can be destroyed by such defects, the occurrence of a phase such as the $L_x$ reported here has not been theoretically predicted. All the theoretical investigations on the sponge phase hitherto have been confined to the steric stabilized regime of the bilayers. It would be interesting to see if the structure presented here can be accounted for by extending the existing models of the sponge phase to the attraction dominated regime.

We thank M. E. Cates, Y. Hatwalne, R. Krishnaswamy, S. Ramaswamy and A. K. Sood for valuable discussions.


\begin{thebibliography}{99}
\bibitem{foot} Our observations differ in two aspects from the predictions in refs. \cite{helfrich78,nelson81,holyst94}. Firstly, the theories predict a nematic phase in the place of the isotropic phase observed here. Secondly, we see a range of coexistence, although the defect mediated transition is expected to be continuous. However, the transition temperature is expected to depend upon the film thickness \cite{holyst94}, which in the presnt case would correspond to the size of the MLV. Therefore, polydispersity in the MLV size might also result in a range of coexistence of the two phases. Further work is needed to confirm if this is indeed the case.
\end{thebibliography}
\end{document}


\section{Supplementary material}

\subsection{Swelling behavior of the $L_\alpha$ phase}

\begin{figure}[h]
\begin{center}
\includegraphics[width=80mm,angle=0] {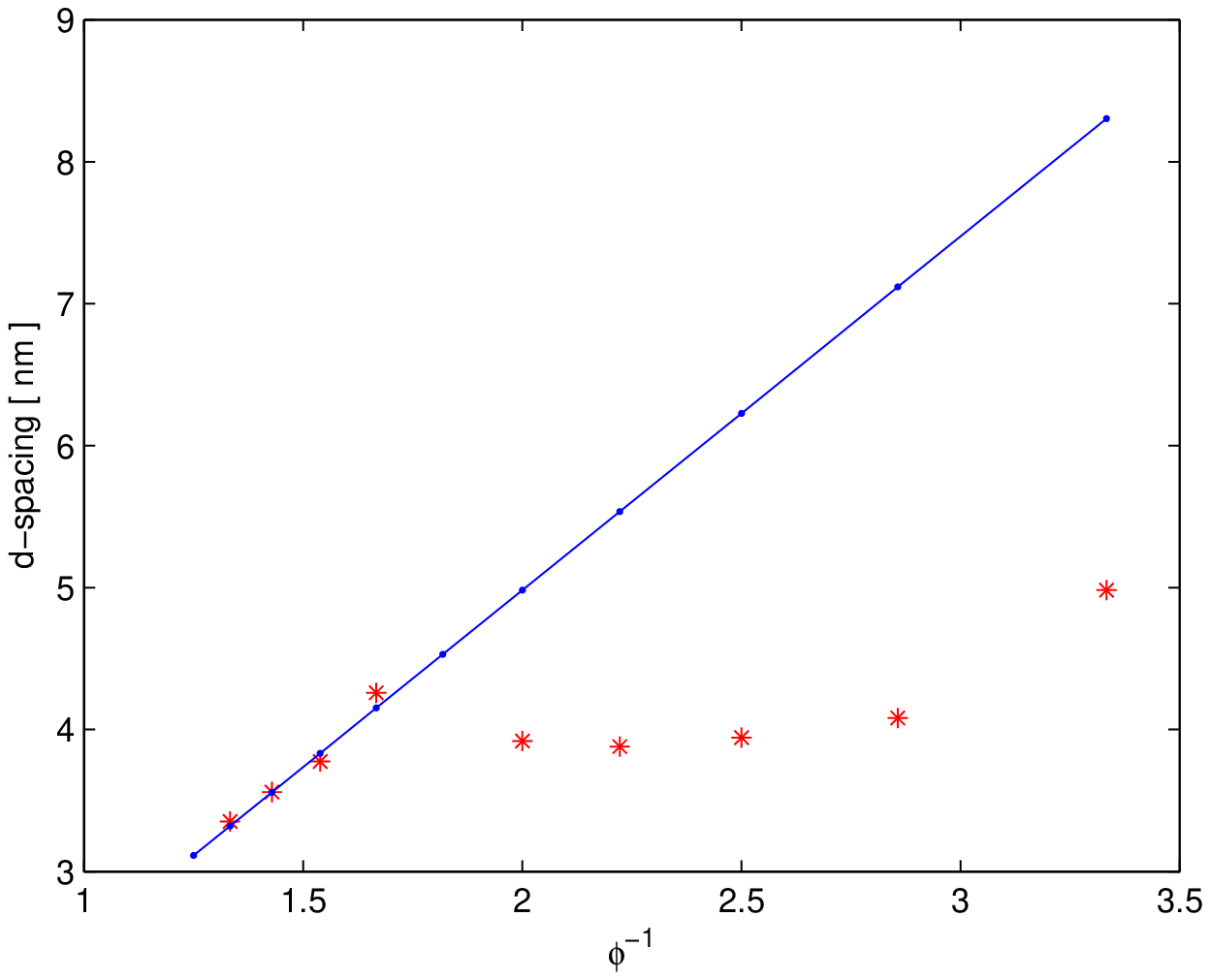}
\end{center}
\caption{Swelling behavior of the lamellar phase of the SDS-PTHC-water system at T =  25 $^\circ$C and $\alpha$ = 1.}
\label{d-swell}
\end{figure}

The variation of the lamellar periodicity with $\Phi_s$ in the $L_\alpha$ phase is shown in  figure \ref{d-swell}.
The maximum water content of the $L_\alpha$ phase is found to be about 35 wt\% beyond which it coexists with excess water forming a MLV dispersion. Within the narrow composition range where the pure $L_\alpha$ phase exists, variation of the lamellar periodicity (d) follows the expected swelling behaviour given  by $ d = \delta /\phi $, where $\delta$ is the bilayer thickness and $\phi$ the amphiphile volume fraction, with $\delta$ = 2.49 nm. Here we have assumed the density of the bilayer to be 1, so that $\phi$ = $\Phi_s$.

\subsection{Analysis of the SAXS data}

The SAXS patterns were analyzed following the procedure described in ref. \cite{pabst00}. The scattered intensity is given by 
\begin{equation}
I(q)= S(q)|F(q)|^2 /q^2 + N_u|F(q)|^2 /q^2,
\label{Int_expr}
\end{equation}
 where S(q) is the structure factor and F(q) the form factor of the bilayers. F(q) is given by the Fourier transform of the bilayer electron density profile $\rho(z)$. $\rho(z)$ is modeled as consisting of two Gaussians of width $\sigma_h$ representing the 
headgroups, centered at z = $\pm$ $z_c$, and a third Gaussian of width $\sigma_c$ at the bilayer center(z = 0) representing 
the terminal methyl groups, 
\begin{equation}
\rho(z) =  \rho_{CH_2}+\bar{\rho}_h[exp(-\frac{(z-z_h)^2}{2\sigma_h^2}+exp(-\frac{(z+z_h)^2}{2\sigma_h^2})  + \bar{\rho}_c(exp-\frac{z^2}{2\sigma_c^2})],
\end{equation}
where the electron densities of the headgroup $\bar{\rho}_h$ and hydrocarbon tails $\bar{\rho}_c$ are defined relative to the methylene electron density $\rho_{CH_2}$.

The last term in eq. 1 is due to diffuse scattering from uncorrelated bilayers. 

The structure factor of the lamellar stack is given by ~\cite{pabst00, zhang94},
\begin{equation}
S(q) = N + 2 \sum^{N-1}_{k=1} (N-k) cos (kqd) \times e^{-(d/2\pi)^2q^2 \eta \gamma} (\pi k)^{-(d/2 \pi)^2 q^2 \eta}
\end{equation}
\noindent
N is the mean number of coherent scattering bilayers in the stack  and $\gamma$ the Eulers' constant. The Caille parameter $\eta = q^2k_BT/(8\pi\sqrt{KB})$, where $k_B$ is the Boltzmann constant, $T$ the temperature, K and B the bending and bulk moduli of the lamellar phase.\par

Although the $L_x$ phase is not a lamellar phase, we have used the above procedure to analyze the SAXS data from this phase, since the SAXS patterns are qualitatively similar to those of the $L_\alpha$ phase, but with a much larger peak width.

Table 1 shows the values of the model parameters for the different phases obtained from the best fit of the calculated and observed SAXS patterns. The calculated SAXS patterns are shown as solid lines in the text and the electron density profiles as insets in the corresponding figures. In the $L_\alpha$ phase the number of correlated bilayers is  $\sim$ 10 whereas in the $L_x$ phase it is  $\sim$ 2. For UB only the form factor of the bilayer is enough to fit the data. Note that values of all other model parameters corresponding to the bilayer electron density profile are comparable in the three phases. Thus the SAXS data show that the $L_x$ phase is made up of bilayers, just as the $L_\alpha$ phase, but with only short-range positional correlations. The higher value of $\eta$ in the $L_x$ compared to $L_\alpha$ is due to the much broader peak in this phase. However, in this case $\eta$ should be treated just as a fitting parameter and should not be interpreted in terms of thermal fluctuations of the bilayers.

  \begin{table*}[h]
	\centering
\begin{ruledtabular}
	\begin{tabular}[t]{*{4}{c}}

\hline \\
				Phase &  UB  &  $L_\alpha$  & $L_x$ \\ 	
				&  $\alpha=1.0$;$\Phi_s$ = 20  	&  $\alpha=1.4$;$\Phi_s$ = 20  &$\alpha$=1.0;$\Phi_s$ = 30 \\ [1ex]		
													& $T = 60^\circ C$  	   		& $T = 10^\circ C$  & $T$ = 90 $^\circ$C	\\
\hline \\
											$z_{H} \ (nm)$ 				& 0.89 $\pm$ 0.06   			 & 0.91 $\pm$ 0.05  	 & 1.02 $\pm$ 0.05  	\\ [1ex]
						$\sigma_{H} \  (nm)$    				&	0.48 $\pm$ 0.04  			   & 0.45 $\pm$ 0.00     &  0.48 $\pm$  0.00  	\\  [1ex]
						$\sigma_{C} \  (nm)$ 				    & 0.86 $\pm$ 0.01   			 & 0.78 $\pm$ 0.1      & 0.46 $\pm$ 0.06  	\\  [1ex]
		$\overline{\rho_{C}} / \overline{\rho_{H}}$				& -1.1	$\pm$ 0.1  						& -1.0 $\pm$ 0.1  	 & -1.95 $\pm$ 0.23	\\ [1ex]
									$d  \ (nm)$						    & $-$									     & 3.98 $\pm$ 0.04  	 & 4.32 $\pm$ 0.04  	\\  [1ex]		
									$\eta $						    & $-$										   & 0.32 $\pm$ 0.02     & 1.31 $\pm$ 0.06		\\  [1ex]				
												N 					        & $-$					 						 & 15 $\pm$ 3 	       & 2 $\pm$ 0	\\  [1ex]								
									$N_u$ 					      & $-$							         & 0	                 & 0.11 $\pm$ 0.15	\\  [1ex]	
					
\hline
	
	\end{tabular}
	\caption{Values of the model parameters obtained for the different phases.}
\end{ruledtabular}
		\end{table*}